\documentclass[prb,aps,twocolumn,draft,tightenlines,showpacs,floatfix]{revtex4}
\usepackage{psfig}
\usepackage{amssymb}
\begin{document}
\title{Spin dephasing in $n$-typed GaAs quantum wells}
\author{M. Q. Weng}%
\author{M. W. Wu}%
\thanks{Author to whom correspondence should be addressed}%
\email{mwwu@ustc.edu.cn}%
\affiliation{Structure Research Laboratory, University of Science \&%
Technology of China, Academia Sinica,  Hefei, Anhui, 230026, China\\%
Department of Physics, University of Science \&%
Technology of China, Hefei, Anhui, 230026, China}%
\altaffiliation{Mailing Address.}
\date{\today}
\begin{abstract}
We perform a many-body study of the spin dephasing 
due to the D'yakonov-Perel' effect 
in $n$-typed GaAs (100) quantum wells for high temperatures ($\geq
120$~K) 
under moderate magnetic fields in the Voigt configuration 
by constructing and numerically solving the
kinetic Bloch equations. We include all the spin conserving scattering 
such as the electron-phonon, the electron-nonmagnetic impurity as
well as  the electron-electron Coulomb scattering in our
theory and investigate how the spin dephasing rate is  
affected by the initial spin polarization,
temperature, impurity, magnetic field as well as the electron density.
The dephasing obtained from our theory
contains not only that due to the  effective spin-flip scattering
first proposed by  D'yakonov and  Perel' [Zh. Eksp. Teor. Fiz. {\bf
  60}, 1954(1971)[Sov. Phys.-JETP {\bf 38}, 1053(1971)]], 
but also the recently proposed many-body dephasing 
due to the inhomogeneous broadening provided by the DP term
[Wu, J. Supercond.:Incorp. Novel Mechanism {\bf 14}, 
245 (2001); Wu and Ning, 
Eur. Phys. J. B {\bf 18}, 373 (2000)]. We show that for the
electron densities we study, the spin dephasing rate
is dominated by the 
many-body effect. Equally remarkable is that 
we are now able to investigate the spin dephasing with extra large
spin polarization (up to 100~\%) which has not been discussed both
theoretically and experimentally. We
find a dramatic decrease of the spin dephasing rate 
for large spin polarizations. 
The spin dephasing time (SDT), which is defined as the inverse of the
spin dephasing rate, we get at low initial spin polarization is in agreement 
with the experiment both qualitatively and quantitatively. 
\end{abstract}

\maketitle

\section {Introduction}

The resent development of ultrafast nonlinear optical 
experiments\cite{damen,wagner,baumberg_1994_prl,baumberg_1994_prb,%
heberle,buss1,crooker_1996,crooker_1997,buss2,kikkawa1,kikkawa2,%
kikkawa3,ohno1,ohno} has stimulated immense interest in spintronics 
in semiconductors as it shows great potential 
of using the spin degree of freedom of electrons in place of/in addition to
the charge degree of freedom for device application such as qubits,
quantum memory devices, and spin transistors. 

In order to make use of the spin degree of freedom in semiconductor
spintronics, it is crucial to have a thorough understanding 
of spin dephasing mechanism.
Three spin dephasing mechanisms have been proposed 
in semiconductors:\cite{meier,aronov}
the Ellit-Yafet (EY) mechanism,\cite{yafet,elliot} the D'yakonov-Perel'
(DP) mechanism,\cite{dp} and the Bir-Aronov-Pikus (BAP)
mechanism.\cite{bap} In the EY mechanism, the spin-orbit interaction leads
to mixing of wave functions of opposite spins. This mixing results in
a nonzero electron spin flip due to impurity and phonon scattering. The DP
mechanism is due to the spin-orbit interaction in crystals without
inversion center, which results in spin state splitting of the
conduction band at $k\not=  0$. This is equivalent to an effective
magnetic field acting on the spin, with its magnitude and orientation
depending on ${\bf k}$. Finally, the BAP mechanism is originated from
the mixing of heavy hole and light hole bands induces by spin-orbit
coupling. Spin-flip (SF) scattering of electrons by holes due to the
Coulomb 
interaction is therefore permitted, which gives rise to spin
dephasing. The dephasing rates of these mechanisms 
for low polarized system are calculated in
the framework of single particle approximation.\cite{meier} 
For GaAs, the EY mechanism is less effective
under most conditions, due to the large band gap and low scattering
rate for high quality samples. The BAP mechanism is important
for either $p$-doped or insulating GaAs. For $n$-doped samples,
however, as holes are rapidly recombined with electrons due to
the presence of a large number of electrons, spin
dephasing due to the regular BAP mechanism is blocked. Therefore,
the DP mechanism  (or possibly the EY mechanism under certain
conditions) is the main mechanism of spin dephasing for $n$-type GaAs.

It is important to note that all the mechanisms above either provide 
or are treated as SF scattering.  
Spin-conserving (SC) scattering,  such as  the ordinary  
Coulomb scattering,
 electron-phonon  and  electron-nonmagnetic 
impurity  scattering which has been extensively studied in 
connection with optical dephasing and relaxation,\cite{haug}
are commonly believed to be unable to cause spin dephasing 
as the corresponding 
interaction Hamiltonians commute with the total spin operator. 

Recently Wu proposed a many-body kinetic theory\cite{wu_js_2001} to 
study the spin precession and spin dephasing
in insulating ZnSe/Zn$_{1-x}$Cd$_x$Se quantum well (QW),\cite{wu_prb_2000}  
$n$-typed bulk GaAs samples,\cite{wu_pss_2000} and $n$-typed GaAs (110)
QW,\cite{wu_ssc_2002} 
under moderate magnetic fields in the Voigt configuration.
Based on this many-body theory, he further showed that the SC scattering
can also cause spin dephasing in the presence of inhomogeneous
broadening.\cite{wu_pss_2000,wu_ssc_2002,wu_jpsj_2001,%
wu_epjb_2000,wu_js_2001} This 
novel spin dephasing mechanism 
has long been overlooked in
the literature. Differing from the earlier study of the spin dephasing
which comes from SF scattering, the spin dephasing through
inhomogeneous broadening is caused by irreversibly disrupting the
phases between spin dipoles and is therefore a many-body
effect.\cite{wu_jpsj_2001,wu_epjb_2000,wu_js_2001} 
Very recently we have shown that this inhomogeneous broadening effect
also plays an important role in the spin 
transport\cite{weng_prb_2002,weng_jap_2003}.

In this paper, we study the spin dephasing in
$n$-doped GaAs (100) QW's. We  calculate the SDT by numerically
solving the many-body kinetic equations with all the scattering included.
Differing from the previous investigation in the bulk case where
we are only able to get the SDT qualitatively,\cite{wu_pss_2000} 
here we  get the SDT quantitatively thanks to the
reduction of dimension in the momentum space.
Moreover as we include the electron-electron
Coulomb scattering in our calculation, 
for the first time we are  able to study the spin dephasing
with extra large  initial spin polarization (up to 100\ \%) which has
not been investigate both experimentally and theoretically. 

We organize the paper as follows: We present our model and the kinetic
equations in Sec.\ II. Then in 
Sec.\ III(A) we show the time evolution of the spin signal where we
show the contribution of the Coulomb scattering to the spin dephasing.
In Sec.\ III(B) we investigate how the SDT changes with the variation
of the initial spin polarization. The temperature dependence of the
SDT under different spin polarization is discussed in detail 
in Sec.\ III(C), where we also highlight
the difference between the present many-body theory and the earlier
simplified theory. In Sec.\ III(D) we show the magnetic field 
dependence of the SDT. Finally we discuss how the electron density
affect the SDT. We present the conclusion and summary in Sec.\ IV.

\section {Kinetic Equations}

We start our investigation from an $n$-doped (100) GaAs QW with 
well width $a$. The growth direction is assumed to be $z$-axis. 
A moderate magnetic field {\bf B} is applied along the $x$ axis. 
Due to the confinement of the QW, the
momentum states along $z$ axis are quantized.  Therefore the electron
states are characterized by a subband index $n$ and a two dimensional
wave vector ${\bf k}=(k_x, k_y)$ together with a spin index $\sigma$.
In the present paper, the subband separation is assumed to be large
enough so that only the lowest subband is populated and the transition
to the upper subbands is unimportant. Therefore, one only needs to
consider the lowest subband. For $n$-doped samples, spin dephasing
mainly comes from the DP mechanism.\cite{dp} With the DP term
included, the Hamiltonian of the electrons in the QW takes the form:
\begin{equation}
  H=\sum_{{\bf k}\sigma\sigma^{\prime}}\biggl\{
\varepsilon_{\bf k}+\bigl[g\mu_B{\bf B}+{\bf h}({\bf k})\bigr]
\cdot{\vec{\bf \sigma}_{\sigma\sigma^{\prime}}\over 2}\biggr\}
c^{\dagger}_{{\bf k}\sigma}c_{{\bf k}\sigma^{\prime}}+H_I.
\label{eq:hamiltonian}
\end{equation}
Here $\varepsilon_{{\bf k}}={\bf k}^2/2m^{\ast}$ is the energy of
electron with wavevector ${\bf k}$ and effective mass $m^{\ast}$.
$\vec{\bf \sigma}$ are the Pauli matrices. 
In QW system, the DP term is composed of the Dresselhaus
term\cite{dress} and the Rashba term.\cite{ras,rashba} The Dresselhaus
term is due to the lack of inversion symmetry in the zinc-blende
crystal Brillouin zone and is sometimes referred to as bulk inversion
asymmetry (BIA) term.  For the (100) GaAs QW system, it can be written
as\cite{eppen,ivch}
\begin{eqnarray}
  &&h^{\mbox{BIA}}_x({\bf k})=\gamma k_x(k_y^2-\langle k_z^2\rangle), \;
  \nonumber\\
  &&h^{\mbox{BIA}}_y({\bf k})=\gamma k_y(\langle k_z^2\rangle-k_x^2), \;
\nonumber\\
&&  h^{\mbox{BIA}}_z({\bf k})=0\ .
  \label{eq:dp}
\end{eqnarray}
Here $\langle k^2_z\rangle$ represents the average of the operator
$-({\partial\over\partial z})^2$ over the electronic state of the
lowest subband and is therefore $(\pi/a)^2$.
$\gamma=(4/3)(m^{\ast}/m_{cv})(1/\sqrt{2m^{\ast
3}E_g})(\eta/\sqrt{1-\eta/3})$ and $\eta=\Delta/(E_g+\Delta)$, in
which $E_g$ denotes the band gap; $\Delta$ represents the spin-orbit
splitting of the valence band; $m^{\ast}$ standing for the electron mass
in GaAs; and $m_{cv}$ is a constant close in magnitude to free
electron mass $m_0$.\cite{aronov} 
Whereas the Rashba term appears if the
self-consistent potential within a QW is asymmetric along the growth
direction and is therefore referred to as structure inversion
asymmetry (SIA) contribution. Its scale 
is proportional to the interface electric
field along the growth direction. For narrow band-gap
semiconductors such as InAs, the Rashba term is the main
spin-dephasing mechanism; whereas in the wide band-gap semiconductors
such as GaAs, the Dresselhaus term is dominant. 
In the present paper, we will take only the Dresselhaus term into
consideration as we focus on the spin dephasing in GaAs QW. 
The interaction Hamiltonian $H_I$ is composed of Coulomb interaction
$H_{ee}$, electron-phonon interaction $H_{ph}$, as well as
electron-impurity scattering $H_i$. Their expressions can be found in
textbooks.\cite{haug,mahan} 

We construct the kinetic Bloch equations by the nonequilibrium Green
function method\cite{haug} as follows:
\begin{equation}
  \label{eq:bloch}
  \dot{\rho}_{{\bf k},\sigma\sigma^{\prime}}
  =\dot{\rho}_{{\bf k},\sigma\sigma^{\prime}}|_{\mbox{coh}}
  +\dot{\rho}_{{\bf k},\sigma\sigma^{\prime}}|_{\mbox{scatt}}
\end{equation}
Here $\rho_{{\bf k}}$ represents the single particle density
matrix. The diagonal elements describe the electron distribution
functions $\rho_{{\bf k},\sigma\sigma}=f_{{\bf k}\sigma}$. The
off-diagonal elements $\rho_{{\bf k},{1\over
    2}-{1\over2}}\equiv\rho_{{\bf k}}$ describe 
the inter-spin-band polarizations
(coherence) of the spin coherence.\cite{wu_prb_2000} Note that 
$\rho_{{\bf k},-{1\over 2}{1\over 2}}\equiv \rho^{\ast}_{{\bf k},{1\over
    2}-{1\over 2}}=\rho^{\ast}_{{\bf k}}$. Therefore, $f_{{\bf
    k}\pm{1\over 2}}$ and $\rho_{{\bf k}}$ are the quantities to be
determined from Bloch equations. 

The coherent part of the equation of motion for the electron
distribution function is given by 
\begin{widetext}
\begin{equation}
  \label{eq:f_coh}
  {\partial f_{{\bf k},\sigma}\over \partial t}|_{\mbox{coh}}=
-2\sigma\bigl\{[g\mu_BB+h_x({\bf k})]\mbox{Im}\rho_{{\bf k}}+h_y({\bf k})
\mbox{Re}\rho_{{\bf k}}\bigr\}
+4\sigma\mbox{Im}\sum_{{\bf q}}V_{{\bf q}}\rho^{\ast}_{{\bf k}+{\bf
    q}} \rho_{{\bf k}},
\end{equation}
where $V_{{\bf q}}=4\pi e^2/[\kappa_0(q+q_0)]$ is the 2D Coulomb
matrix element under static screening. 
$q_0=(e^2m^{\ast}/\kappa_0)\sum_{\sigma}f_{{\bf k}=0,\sigma}$  and
$\kappa_0$ is the static dielectric constant. 
The first term
on the right hand side (RHS) of Eq.~(\ref{eq:f_coh}) describes spin
precession of electrons under the magnetic field ${\bf B}$ as well as
the effective magnetic field ${\bf h}({\bf k})$ due to the DP
effect. The scattering terms of electron distribution functions in the
Markovian limit are given by 
\begin{eqnarray}
  \label{eq:f_scatt}
  {\partial f_{{\bf k},\sigma} \over \partial t}|_{\mbox{scatt}} &=& 
  \biggl\{-2\pi\sum_{{\bf q}q_z\lambda}g_{{\bf Q}\lambda}^2
  \delta(\varepsilon_{{\bf k}}-\varepsilon_{{\bf k}-{\bf
      q}}-\Omega_{{\bf q}q_z\lambda})
  \bigl[N_{{\bf q}q_z\lambda}
  (f_{{\bf k}\sigma}-f_{{\bf k}-{\bf q}\sigma})
  +f_{{\bf k}\sigma}(1-f_{{\bf k}-{\bf q}\sigma})
  -\mbox{Re}(\rho_{{\bf k}}\rho^{\ast}_{{\bf k}-{\bf q}})\bigr]
  \nonumber\\
 && -2\pi N_i\sum_{{\bf q}}U^2_{{\bf q}}
  \delta(\varepsilon_{{\bf k}}-\varepsilon_{{\bf k}-{\bf q}})
  \bigl[f_{{\bf k}\sigma}(1-f_{{\bf k}-{\bf q}\sigma})-
  \mbox{Re}(\rho_{{\bf k}}\rho^{\ast}_{{\bf k}-{\bf q}})\bigr]
  -2\pi\sum_{{\bf q}{\bf k}^{\prime}\sigma^{\prime}}V_{{\bf q}}^2 
  \delta(\varepsilon_{{\bf k}-{\bf q}}-\varepsilon_{{\bf k}}
  +\varepsilon_{{\bf k}^{\prime}}-
  \varepsilon_{{\bf k}^{\prime}-{\bf q}})
  \nonumber\\
  &&\Bigl[
  (1-f_{{\bf k}-{\bf q}\sigma})f_{{\bf k}\sigma}
  (1-f_{{\bf k}^{\prime}\sigma^{\prime}})
  f_{{\bf k}^{\prime}-{\bf q}\sigma^{\prime}}
  +{1\over 2}\rho_{{\bf k}}\rho^{\ast}_{{\bf k}-{\bf q}}
  (f_{{\bf k}^{\prime}\sigma^{\prime}}-
  f_{{\bf k}^{\prime}-{\bf q}\sigma^{\prime}})
  +{1\over 2}\rho_{{\bf k}^{\prime}}
  \rho^{\ast}_{{\bf k}^{\prime}-{\bf q}}
  (f_{{\bf k}-{\bf q}\sigma}-f_{{\bf k}\sigma})\Bigr]
  \biggr\}\nonumber\\
  &&-\{{\bf k}\leftrightarrow{\bf k}-{\bf q},{\bf
  k}^{\prime}\leftrightarrow{\bf k}^{\prime}-{\bf q}\},
\end{eqnarray}
\end{widetext}
\noindent in which $\{{\bf k}\leftrightarrow{\bf k}-{\bf q},
{\bf k}^{\prime}\leftrightarrow{\bf k}^{\prime}-{\bf q}\}$ stands for the
same terms as in the previous $\{\}$ but with the interchange ${\bf
  k}\leftrightarrow {\bf k}-{\bf q}$ and ${\bf
  k}^{\prime}\leftrightarrow{\bf k}^{\prime}-{\bf q}$. 
The first term inside the braces on the RHS of Eq.~(\ref{eq:f_scatt})
comes from the electron-phonon interaction. $\lambda$ stands for the
different phonon modes, {\em i.e.}, one longitude optical (LO) phonon mode,
one longitudinal acoustic (AC) phonon mode 
due to the deformation potential, and two AC modes due to the
transverse piezoelectric field. $g_{{\bf q}q_z\lambda}$ are the matrix
elements of electron-phonon coupling for mode $\lambda$. 
For LO phonons, $g^2_{{\bf q}q_z \mbox{LO}}=
\{4\pi\alpha\Omega_{\mbox{LO}}^{3/2}/[\sqrt{2\mu}(q^2+q_z^2)]\}
|I(iq_z)|^2$ with 
$\alpha=e^2\sqrt{\mu/(2\Omega_{\mbox{LO}})}
(\kappa^{-1}_{\infty}-\kappa_0^{-1})$. $\kappa_{\infty}$ is the
optical dielectric constant and $\Omega_{\mbox{LO}}$ is the LO phonon
frequency. The form factor 
$|I(iq_z)|^2=\pi^2\sin^2y/[y^2(y^2-\pi^2)^2]$ with $y=q_za/2$. 
$N_{{\bf q}q_z\lambda}=1/[\exp(\Omega_{{\bf q}q_z\lambda}/k_BT)-1]$ is
the Bose distribution of phonon mode $\lambda$ at temperature $T$. 
The second term inside the braces on the RHS of Eq.~(\ref{eq:f_scatt})
results from the electron-impurity scattering under the random phase
approximation with $N_i$ denoting the impurity
concentration. $U_{\bf q}^2=\sum_{q_z}\bigl\{4\pi
Z_i e^2/[\kappa_0 (q^2+q_z^2)]\bigr\}^2 |I(iq_z)|^2$ is the
electron-impurity interaction matrix element with $Z_i$ stands for the
charge number of the impurity. $Z_i$ is assumed to be $1$ throughout
our calculation. 
The third term is the contribution of the Coulomb interaction. 
Similarly, the coherent and the scattering parts of the spin coherence
are given by
\begin{widetext}
\begin{equation}
  \label{eq:rho_coh}
  {\partial \rho_{{\bf k}}\over \partial t}\left |_{\mbox{coh}}\right. =
  {1\over 2}[ig\mu_B B + ih_x({\bf k}) + h_y({\bf k})]
  (f_{{\bf k}{1\over 2}}-f_{{\bf k}-{1\over 2}})
  +i\sum_{{\bf q}}V_{\bf q}\bigl[(f_{{\bf k}+{\bf q}{1\over 2}}
  -f_{{\bf k}+{\bf q}-{1\over 2}})\rho_{{\bf k}}
  -\rho_{{\bf k}+{\bf q}}(f_{{\bf k}{1\over 2}}
  -f_{{\bf k}-{1\over 2}})\bigr],
\end{equation}
\begin{eqnarray}
  \label{eq:rho_scatt}
  {\partial \rho_{{\bf k}}\over \partial t}\left |_{\mbox{scatt}}
    \right . &=&\biggl\{
  \pi\sum_{{\bf q}q_z\lambda}g^2_{{\bf q}q_z\lambda}
  \delta(\varepsilon_{{\bf k}}-\varepsilon_{{\bf k}-{\bf q}}
  -\Omega_{{\bf q}q_z\lambda})
  \bigl[\rho_{{\bf k}-{\bf q}}
  (f_{{\bf k}{1\over 2}}+f_{{\bf k}-{1\over 2}})
  +(f_{{\bf k}-{\bf q}{1\over 2}}+f_{{\bf k}-{\bf q}-{1\over 2}}-2)
  \rho_{{\bf k}}
  -2N_{{\bf q}q_z\lambda}(\rho_{{\bf k}}-\rho_{{\bf k}-{\bf q}})\bigr]
  \nonumber \\
  && + \pi N_i\sum_{{\bf q}}U_{{\bf q}}^2
  \delta(\varepsilon_{{\bf k}}-\varepsilon_{{\bf k}-{\bf q}})
  \bigl[(f_{{\bf k}{1\over 2}}+f_{{\bf k}-{1\over 2}})
  \rho_{{\bf k}-{\bf q}}
  -(2-f_{{\bf k}-{\bf q}{1\over 2}}-f_{{\bf k}-{\bf q}-{1\over 2}})
  \rho_{{\bf k}}\bigr]\nonumber\\
  &&-\sum_{{\bf q}{\bf k}^{\prime}}\pi V_{{\bf q}}^2
\delta(\varepsilon_{{\bf k}-{\bf q}}-\varepsilon_{{\bf
    k}}+\varepsilon_{{\bf k}^{\prime}}-\varepsilon_{{\bf
    k}^{\prime}-{\bf q}})
\biggl(
\bigl(f_{{\bf k}-{\bf q}{1\over 2}}\rho_{{\bf k}}
+\rho_{{\bf k}-{\bf q}}f_{{\bf k}-{1\over 2}}
\bigr)
(f_{{\bf k}^{\prime}{1\over 2}}-
f_{{\bf k}^{\prime}-{\bf q}{1\over 2}}
+f_{{\bf k}^{\prime}-{1\over 2}}-
f_{{\bf k}^{\prime}-{\bf q}-{1\over 2}})\nonumber\\
&&+\rho_{{\bf k}}\bigl[
(1-f_{{\bf k}^{\prime}{1\over 2}})f_{{\bf k}-{\bf q}{1\over 2}}
+(1-f_{{\bf k}^{\prime}-{1\over 2}})f_{{\bf k}-{\bf q}-{1\over 2}}
-2\mbox{Re}(\rho^{\ast}_{{\bf k}^{\prime}}\rho_{{\bf k}-{\bf q}})
\bigr]
- \rho_{{\bf k}-{\bf q}}\bigl[
f_{{\bf k}^{\prime}{1\over 2}}(1-f_{{\bf k}^{\prime}-{\bf q}{1\over 2}})
\nonumber\\ &&
+(1-f_{{\bf k}^{\prime}-{1\over 2}})f_{{\bf k}^{\prime}-{\bf q}-{1\over 2}}
-2\mbox{Re}(\rho^{\ast}_{{\bf k}^{\prime}}\rho_{{\bf k}^{\prime}-{\bf q}})
\bigr]\biggl)\biggl\}
-\bigl\{{\bf k}\leftrightarrow {\bf k}-{\bf q},{\bf
    k}^{\prime}\leftrightarrow
{\bf k}^{\prime}-{\bf q}\bigr\}\ .
\end{eqnarray}
\end{widetext}
\noindent The initial conditions are taken at $t=0$ as: 
\begin{equation}
\rho_{\bf k}|_{\rm t=0} = 0
\label{eq:rho_init}
\end{equation}
\begin{equation}
f_{{\bf k}\sigma}|_{\rm t=0} = 1/\bigl\{\exp[(\varepsilon_{\bf
  k}-\mu_{\sigma})/k_BT]+1\bigr\} 
\label{eq:fk_init}
\end{equation}
where $\mu_\sigma$ is the chemical potential for spin $\sigma$. The condition
$\mu_{\frac{1}{2}}\neq\mu_{-\frac{1}{2}}$ gives rise to the imbalance
of the electron densities of the two spin bands. Eqs.~(\ref{eq:bloch})
through (\ref{eq:rho_scatt}) together with the initial conditions 
Eqs.~(\ref{eq:rho_init}) and (\ref{eq:fk_init}) comprise the complete
set of kinetic Bloch equations of our investigation.

\section {Numerical Results}
The kinetic Bloch equations form a set of nonlinear equations. All the
unknowns to be solved appear in the scattering terms. Specifically,
the electron distribution function is no longer a Fermi distribution
because of the existence of the anisotropic DP term ${\bf h}({\bf
  k})$. This term in the coherent part drives the electron
distribution away from an isotropic Fermi distribution. The
scattering term attempts to randomize electrons in ${\bf
  k}$-space. Obviously, both the coherent part and the scattering
terms have to be solved self-consistently to obtain the distribution
function and the the spin coherence. 

\begin{table}[htb]
  \centering
  \begin{tabular}{lllllll}
    \hline
    $\kappa_\infty$ & \mbox{}\hspace{1.25cm} &
    10.8 & \mbox{}\hspace{1.25cm} &
    $\kappa_0$ & \mbox{}
    \hspace{1.25cm} & 12.9\\ 
    
    $\omega_0$ & & 35.4~meV & & $m^*$ & &0.067~$m_0$\\
    $\Delta$ & &0.341~eV & &$E_g$ & &1.55~eV\\
    $g$ & &0.44 &&&\\
    \hline
  \end{tabular}
  \caption{Parameters used in the numerical calculations}
  \label{table1}
\end{table}

\begin{figure}[htb]
  \psfig{figure=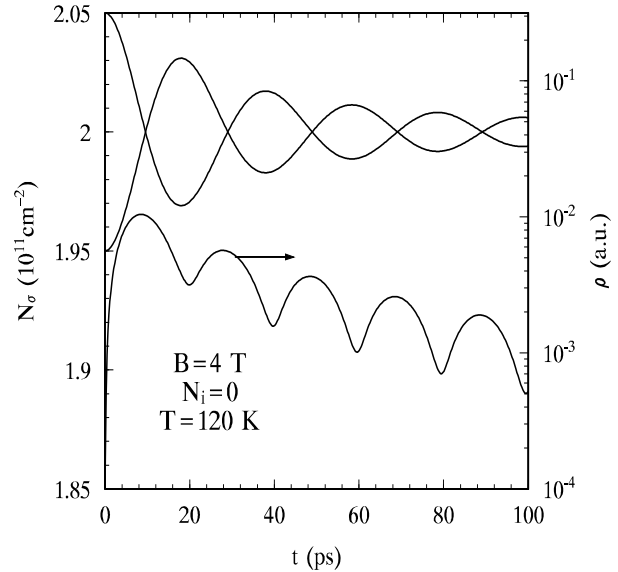,width=9.cm,height=8.5cm,angle=0}
  \caption{Electron densities of up spin and down spin 
    and the incoherently summed spin coherence $\rho$ 
    versus time $t$ without taking account the Coulomb scattering 
    for a GaAs QW with  
    initial spin polarization $P=2.5\%$, 
    at $T=120$~K.
    Note the scale of the spin coherence is on the right side of the
    figure.} 
  \label{fig1}
\end{figure}

We numerically solve the kinetic Bloch equations in such a
self-consistent fashion to study the spin precession between the
spin-up and -down bands. We include electron-phonon scattering
and the electron-electron interaction throughout our computation. 
As we concentrate on the relatively high
temperature regime ($T\geq 120$~K) in the present study, for electron-phonon
scattering we only need to include electron-LO phonon
scattering. Electron-impurity scattering is sometimes excluded. 
As discussed in the previous paper,\cite{wu_prb_2000,kuhn}
irreversible 
spin dephasing can be well defined by the slope of the envelope of the
incoherently 
summed spin coherence $\rho(t)=\sum_{{\bf k}}|\rho_{{\bf k}}|$. 
The material parameters of GaAs for our calculation are
tabulated in Table~\ref{table1}.\cite{made}
The method of the numerical calculation
has been laid out in detail in our previous paper on the
DP mechanism in 3D systems.\cite{wu_pss_2000}
The difference is that here we are able to get the results
quantitatively in stead of only qualitatively as in our previous 3D case,
thanks to the smaller dimension in the momentum space.
Our main results are
plotted in Figs.~\ref{fig1} to \ref{fig11}. 
In these calculations the total electron density $N_e$ 
and the applied magnetic field $B$ are chose to be $4\times
10^{11}$~cm$^{-2}$ and $4$~T respectively unless otherwise specified. 
The width of the quantum well is chosen typical to be $15$~nm except
in the last two figures. 

\subsection {Temporal evolution of the spin signal}

We first study the temporal evolution of the spin signal in a GaAs 
QW at $T=200$~K. 
In Fig.~\ref{fig1} we plot the electron densities in the spin-up and
-down bands together with the incoherently summed spin coherence 
for $N_i=0$. At $t=0$, the initial spin polarization
$P=(N_{1/2}-N_{-1/2})/(N_{1/2}+N_{-1/2})$ is $2.5\%$. 
In this calculation, the Coulomb scattering is not included.
It is seen from the figure that excess electrons in the
spin-up band start to flip to the spin-down band at $t=0$ due to the
presence of the magnetic field and the DP term ${\bf h}({\bf k})$. In
the meantime the spin coherence 
$\rho$ accumulates. At about $9.7$~ps, the electron densities in the
two spin bands become equal and the spin coherence reaches its
maximum. Then the spin coherence starts to feed back and the electron
density in the spin-down band exceeds that in the spin-up band while
$\rho$ deceases. At about $18$~ps, $\rho$ reach its minimum, while
the density difference in the two spin bands reaches its maximum again
with the excess electrons now in the spin-down band. Due to the
the dephasing, the second peak is lower than the first
one. This oscillation goes on until the amplitude of the oscillation
becomes zero due to the dephasing.
\begin{figure}[htb]
  \psfig{figure=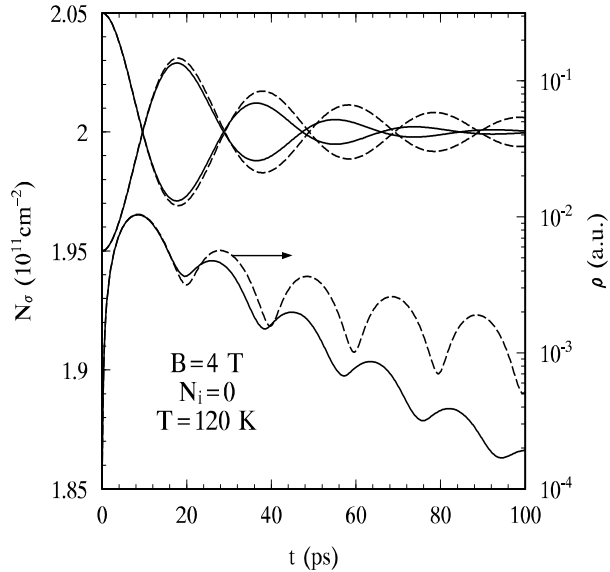,width=9.cm,height=8.5cm,angle=0}
  \caption{Electron densities of up spin and down spin
    and  the incoherently summed spin coherence $\rho$ 
    versus time $t$ with (solid curves) and without (dashed curves)
    taking account the Coulomb scattering  
    for a GaAs QW 
    initial spin polarization $P=2.5\%$ at
    $T=120$\ K.
    Note the scale of the spin coherence is on the right side of the
    figure.} 
  \label{fig2}
\end{figure}

In Fig.~\ref{fig2}, we plot the time evolution of electron densities
in the two spin bands as well as the incoherently summed spin coherence 
for the same GaAs QW system as the previous one but 
taking the Coulomb scattering into account. 
The results without the Coulomb scattering
are replotted as dashed curves in the figure. It is seen from
the figure that for the first oscillation ($t<20$~ps), the
electron densities as well as the spin coherences are almost the same
in the the presence and absence of the Coulomb
scattering. As time goes on, the curves with the Coulomb
scattering deviate from the ones without
the Coulomb scattering. The decay rates of the 
excess spin density as well as the spin coherence 
are faster in the presence of Coulomb scattering. 

It is known that,
the Coulomb scattering is important only when the electron
distribution is divagated from the Fermi function. In the first few
picoseconds, the electrons in the two spin bands flip to their
opposite bands due to the magnetic fields. The buildup of the
inhomogeneity of ${\bf k}$ in the electron distribution function comes
from the DP term
is marginal, and the electron distributions remain approximately
the Fermi function. As time goes on, the effect of the DP term
accumulates, the electron distribution functions divagate
further and further away
from the Fermi function and hence the Coulomb scattering
becomes more and more important. Consequently, the spin signal that
with the Coulomb scattering differs from the one without the Coulomb
scattering. It has been pointed out that the spin dephasing due to the
DP term and the SC scattering comes from two
effects.\cite{wu_pss_2000} The first one is widely discussed that
is due to the anisotropic property of the DP term,
which, combined with SC scattering gives rise to the effective SF
scattering.\cite{meier,dp} In this case, inclusion of additional scattering
enhances the momentum relaxation rate and consequently reduces
the spin dephasing rate.\cite{meier,dp,wu_js_2001}
It has been pointed out that the electron-electron Coulomb
scattering, although does not contribute to the momentum relaxation
rate, trends to reduce the spin dephasing\cite{glazov_2002} based on
the similar 
analysis as in Refs.\ \onlinecite{meier,dp} and \onlinecite{wu_js_2001}.
The second is that the DP term itself also introduces an
inhomogeneous broadening, which in the presence of  the SC scattering
provides {\em additional} spin dephasing
channel\cite{wu_pss_2000,wu_ssc_2002,wu_jpsj_2001,wu_epjb_2000,wu_js_2001}
and therefore results in a faster spin dephasing.
Our calculation self-consistently solves the Kinetic Bloch equations and
includes {\em both} effects. The result
indicates that for the present condition the second effect
is more important and therefore the combined effect by
inclusion of the Coulomb scattering leads to the increase of the
spin dephasing.

\subsection{Spin polarization dependence of the spin dephasing time}

We now turn to study the spin polarization dependence of the SDT.
As our theory is a many-body theory and 
we include all the scattering, especially the
Coulomb scattering, in our calculation, we are able to calculate the 
SDT with large spin polarization. 

\begin{figure}[htb]
  \psfig{figure=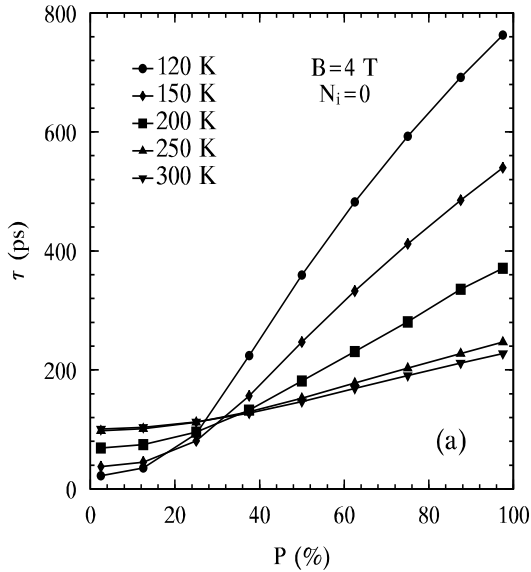,width=9.cm,height=8.5cm,angle=0}
  \psfig{figure=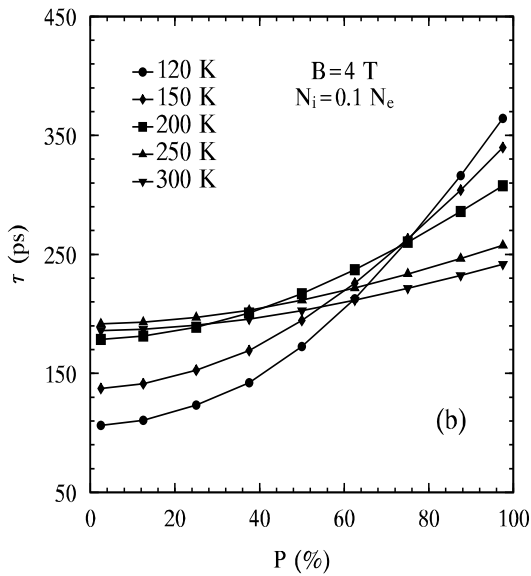,width=9.cm,height=8.5cm,angle=0}
  \caption{Spin dephasing time $\tau$ versus the initial spin
    polarization 
    with different impurity concentration and different temperatures. 
    The impurity densities in (a) and
    (b) are 0 and $0.1 N_e$ respectively. The lines are plotted for
    the aid of eyes. 
  }
  \label{fig3}
\end{figure}

In Fig.~\ref{fig3}, the SDT $\tau$ is
plotted against the initial spin polarization $P$ for GaAs QW's
with $N_i=0$ [Fig.~\ref{fig3}(a)] and $N_i=0.1 N_e$ [Fig.~\ref{fig3}(b)]
at different temperatures. 
The most striking feature of the impurity-free
case is the huge increase of the SDT 
in low temperatures.
For  $T=120$\ K, the SDT increases from 25~ps at low
polarization to 720~ps at $\sim 100$~\% polarization. In other
words, the spin dephasing rate is decreased more than one order  of
magnitude when
the spin polarization increases from 0 to 100~\%. It is also seen from the
figure that the increase of the SDT is reduced with the increase of
temperature. For $T=300$~K, the SDT only
gets an 80~\% increase when the polarization increases from 0 to
100~\%. 

\begin{figure}[htb]
  \psfig{figure=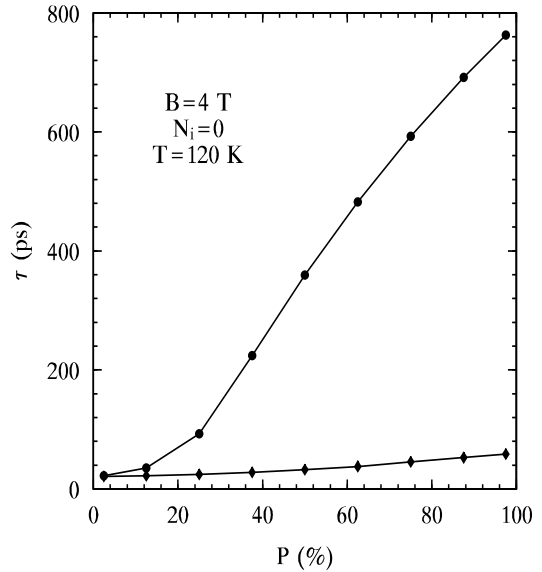,width=9.cm,height=8.5cm,angle=0}
  \caption{Spin dephasing time $\tau$ versus the initial spin
    polarization $P$ for 
    at $T=120$~K and $Ni=0$. 
    Circle ($\bullet$): With the longitudinal component of HF term
    included; 
    Diamond ($\blacklozenge$): Without the longitudinal component of HF term
    include.
    The lines are plotted for
    the aid of eyes. 
  }
  \label{fig4}
\end{figure}

The dramatic increase in the $\tau$-$P$ curve
in the low temperature regime originates from the electron-electron
interaction, specifically the Hartree-Fock (HF) self-energy [{\em i.e.},
the last terms in the Eq. (\ref{eq:f_coh}) 
and (\ref{eq:rho_coh})].  As we know that the  HF term
itself does not contribute to the spin dephasing 
directly.\cite{wu_epjb_2000,wu_js_2001}
However, it behaves as an effective magnetic field which  
can alert the motion of the electron spins and can 
therefore affect the spin dephasing by combining with the DP term. 
For small spin polarization
as commonly discussed in the literature, the
contribution of the HF term is marginal. However, when the polarization
gets higher, the HF contribution becomes larger. 
For example, the magnitude of the effective magnetic field of HF term
is larger than 40~T, which is about ten folds of the applied magnetic
field, for the case when the temperature is 120~K and the spin polarization
is $\sim 100$~\%. Differing from the applied magnetic field which in the Voigt
configuration only gets the
transverse components ({\it i.e.} $B_x$ and $B_y$)  and always causes the
electrons to flip between spin-up and -down bands, the effective
magnetic field formed by the HF term contains a longitudinal component
[$B_z^{\mbox{HF}}({\bf k})$]
which can effectively reduce the ``detuning'' of the spin-up and -down 
electrons, and thus strongly reduces the spin dephasing. In order to
show this detuning effect, we remove the longitudinal component of the effective
magnetic field $B_z^{\mbox{HF}}({\bf k})$ and recalculate the SDT at the
temperature of $120$~K for different initial spin polarization for an
impurity free sample. 
The result is plotted in Fig.~\ref{fig4}. From the figure one
can see that when the longitudinal component of the HF term is removed,
the dramatic increase in the $\tau$-$P$ curve disappears and 
the SDT is insensitive to the initial spin polarization. 

When the temperature increases, for a given initial spin polarization 
the HF term becomes smaller as the electrons are distributed to a wider
range in the ${\bf k}$-space. Therefore the effect of the HF term becomes
smaller too. Consequently the increase in the $\tau$-$P$ curve becomes
slower. 

The $\tau$-$P$ curve gets flatter when the impurities are
introduced. It is shown from Fig.~\ref{fig3}(b) that, when the density
of impurity is large, say $N_i=0.1 N_e$, the fast rise in the $\tau$-$P$
curve in low temperature regime still remains. Nevertheless the rate of 
increase is
much smaller than the corresponding one when the impurities are
absent.
\begin{figure}[htb]
  \psfig{figure=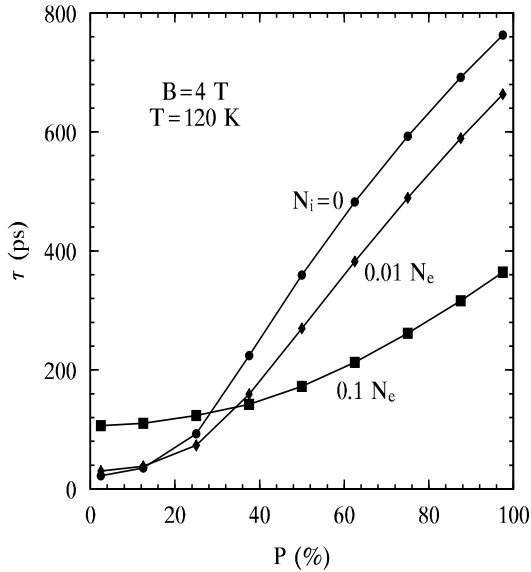,width=9.cm,height=8.5cm,angle=0}
  \caption{Spin dephasing time $\tau$ versus the initial spin
    polarization $P$ at $T=120$~K for 
    different
    impurity levels.  
    Circle ($\bullet$): $N_i=0$;
    Diamond ($\blacklozenge$): $N_i=0.01N_e$;
    Square ($\blacksquare$): $N_i=0.1N_e$. The lines are plotted for
    the aid of eyes. 
  }
  \label{fig5} 
\end{figure}

To further reveal the contribution of the impurity to the dephasing
under different conditions, we plot the  SDT 
as a function of the polarization for different impurity levels at 
$T=120$\ K in Fig.~\ref{fig5}.
The figure clearly shows that when the impurity concentration
increases, the slope of the $\tau$-$P$ curve becomes smaller. This is
because that impurity scattering reduces  the HF term and the
effect of the longitudinal component of the HF term is also
reduced. Consequently, the increase of SDT with the polarization is
reduced.

It is interesting to note that contrary to the high polarization
regime where the SDT decreases with the impurity concentration, the
spin dephasing is reduced by the impurity scattering in the low
polarization regime. 
As we pointed out before that the impurities  affect the spin
dephasing in two ways.\cite{wu_pss_2000}
On the one hand, the electron-impurity
scattering provides a new spin dephasing channel by
combining with the DP term\cite{dp,wu_pss_2000} to give an
effective SF scattering 
through the inhomogeneous broadening
introduced by the DP term.
This gives rise to the enhancement of the dephasing. 
On the other hand, the scattering also redistributes the
electrons in the momentum space and leads them to an isotropic
distribution. Therefore, the scattering can suppress the anisotropy
caused by the DP term, consequently the effective SF scattering.
Moreover, the suppression of the anisotropy also corresponds to the
reduction of the inhomogeneous broadening. Both lead to a smaller spin
dephasing. Our results indicate that in 
the low temperature and the low polarization regime, the impurities
tend to reduce the spin dephasing. 

\subsection{The temperature dependence of the spin dephasing time}

Above we discussed the dependence of spin dephasing on
initial spin polarization for different temperatures. 
Now we turn to the temperature dependence of
the SDT under different initial spin polarizations. 
From Fig.~\ref{fig3}(a) and (b) in Sec.\ III(B),
one can see that for small spin  polarization, 
the SDT increases with the temperature. Whereas in high
polarized regime, the SDT decreases with the temperature. For moderate
polarization, the temperature dependence  is too complicated 
to be described by a monotonic function of temperature. 
Under certain condition, the SDT can be insensitive to the temperature,
e.g. the SDT is almost unchanged with the temperature when the
polarization $P=75$~\% and $N_i=0.1 N_e$. 

\begin{figure}[htb]
  \psfig{figure=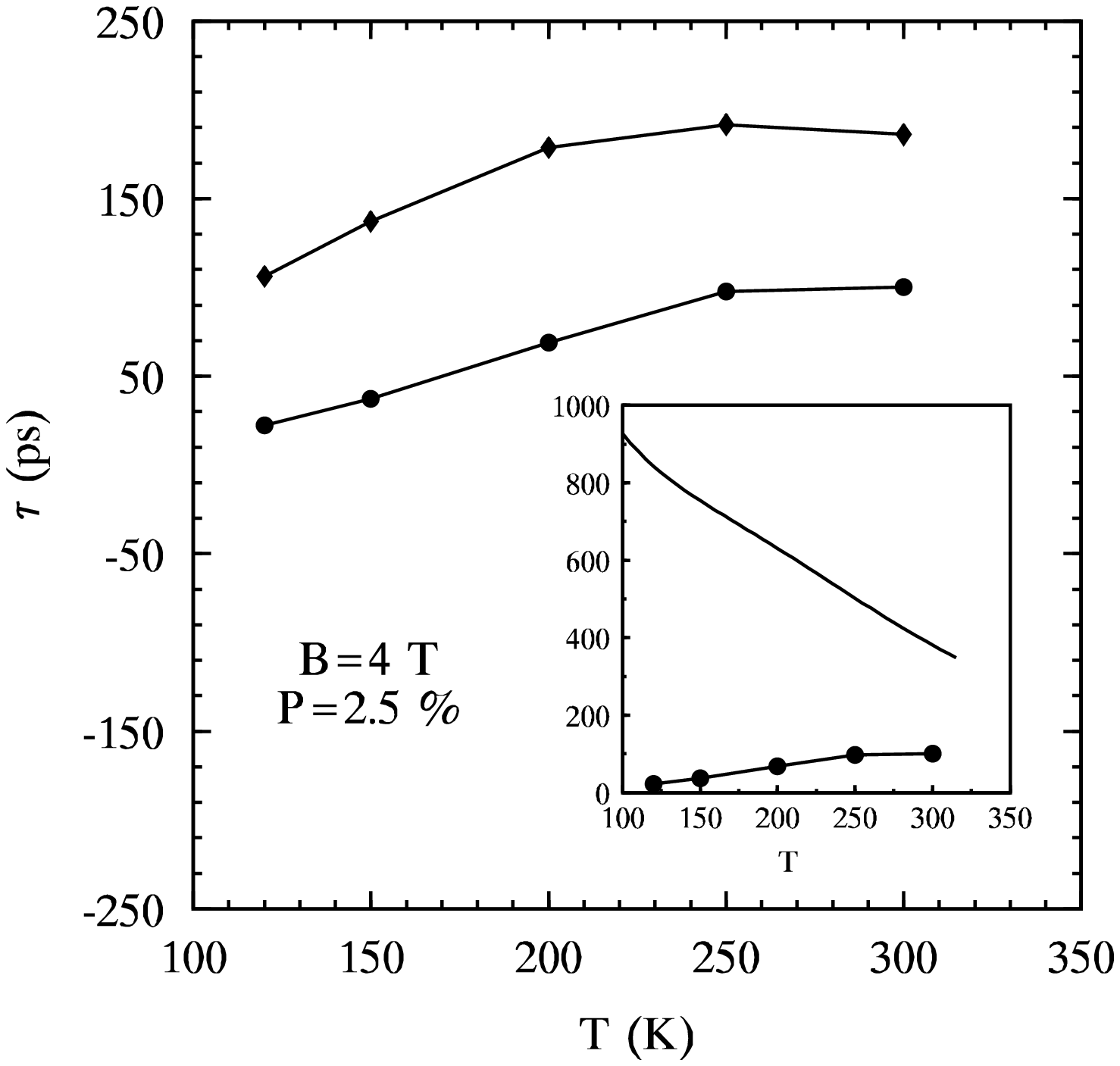,width=9.cm,height=8.5cm,angle=0}
  \psfig{figure=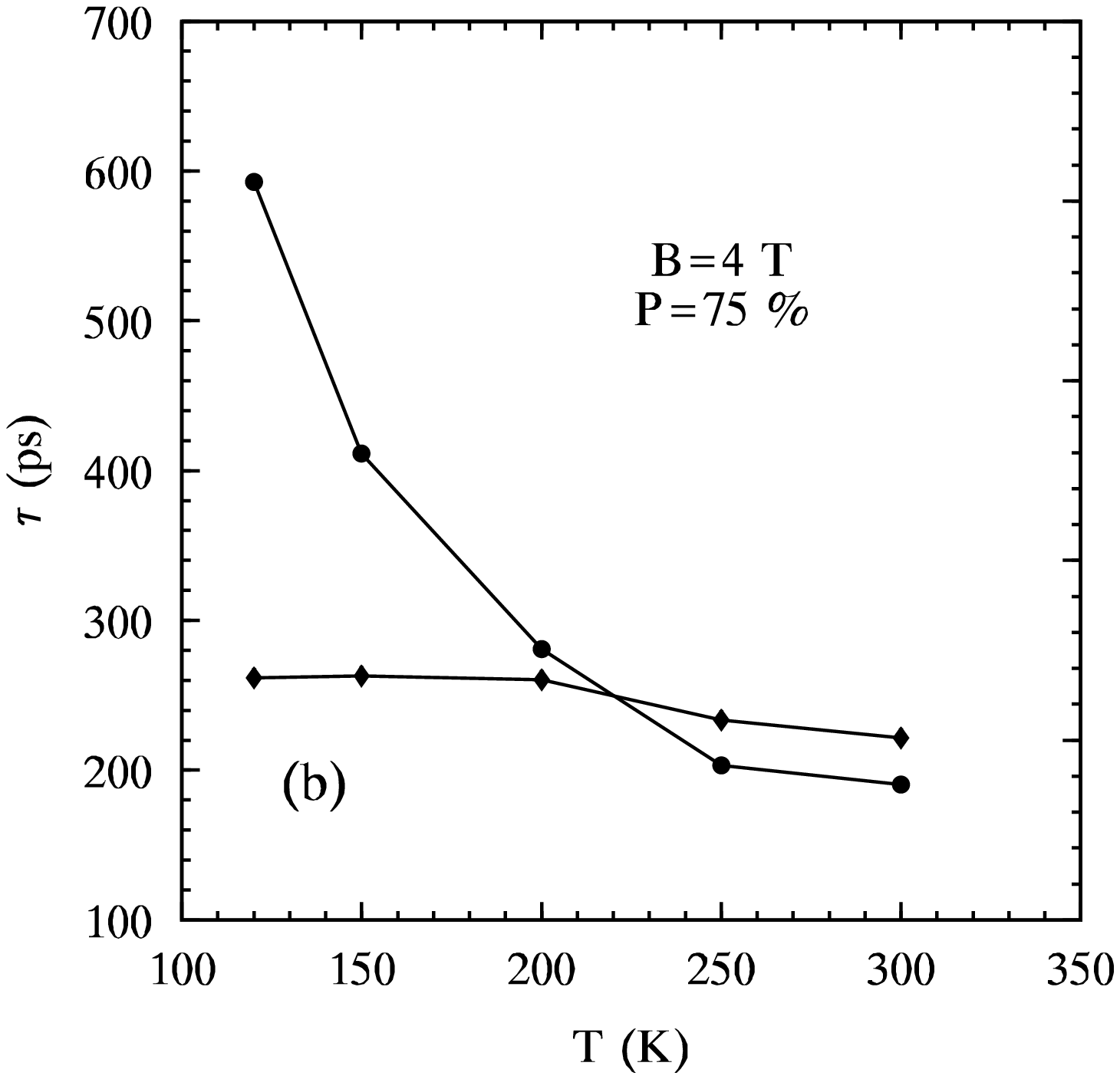,width=9.cm,height=8.5cm,angle=0}
  \caption{Spin dephasing time $\tau$ versus the temperature $T$
    with spin polarization $P=2.5$~\% (a) and $P=75$~\% (b) 
    under two different impurity levels. 
    Circle ($\bullet$): $N_i=0$; Square ($\blacksquare$): $N_i=0.1 N_e$.
    The lines are plotted for eye aid. 
    The SDT predicated by the simplified treatment of DP term (solid
    curve) and our model (circle) for $N_i=0$ is plotted in the inset
    of (a) for comparison. 
}
  \label{fig6}
\end{figure}

To see more detail of how the spin dephasing depends on the
temperature, we replot in Fig.~\ref{fig6}(a) and (b)
the SDT shown in Figs.~\ref{fig3} and \ref{fig4} as a function of
the temperature  
for different impurity levels and different spin polarizations.
It is seen from the figure that, for low polarization, 
the SDT increases systematically with the temperature
for all impurity levels. 
This property is {\em opposite} to the results of earlier simplified
treatments of the DP effect, where it was predicted that the spin
lifetime decreases with the increase of temperature in the 2D
system.\cite{averkiev,dyakonov}
The SDT based on the simplified model is given by\cite{wu_jpsj_2001,%
averkiev,lau}
\begin{equation}
  \label{eq:taus_ani}
  {1\over \tau} = {\int_0^{\infty} d E_{k} \bigl( 
    f_{k{1\over 2}} - f_{{k}-{1\over 2}}\bigr)\Gamma(k)
    \over \int_0^{\infty} d E_{k} \bigl( f_{k{1\over
    2}}-f_{{k}-{1\over 2}}\bigr)}, 
\end{equation}
in which 
\begin{eqnarray}
  \Gamma(k)& =&  {2\tau_{1}(k)}\Bigl[\bigl(\gamma\langle
    k_z^2\rangle\bigr)^2 k^2 - 
    {1\over 2}\gamma\langle k_z^2\rangle k^4 \nonumber\\
&&\mbox{}\hspace{1cm} + {{1+\tau_3(k)/\tau_1(k)}\over 16}\gamma^2 k^6\Bigr]
  \label{eq:gammak}
\end{eqnarray}
and 
\begin{equation}
  \label{eq:tau_n}
  \tau_n^{-1}(k) =
  \int_0^{2\pi}\sigma(E_k,\theta)[1-\cos(n\theta)]d\theta\ .
\end{equation}
$\sigma(E_k,\theta)$ stands for scattering cross-section.
For comparison, we plot the SDT
predicated by the earlier model and 
by our present many-body theory  in the inset of Fig.\ \ref{fig6}(a). 
From the inset one can see
that the SDT  predicated by the earlier model is about one order of
magnitude  larger than the one predicated by our theory. 
In the mean time, the SDT
of the earlier mode drops dramatically with the increase of the
temperature. Nevertheless, in our many-body treatment, it rises slightly
with the temperature.

The recent experiments show that the SDT
in $n$-type quantum wells 
increases slightly with the increase of temperature,\cite{hagele} or
is almost unchanged with the temperature.\cite{mali} 
It is also noted that the SDT's measured in the experiment\cite{mali}
for GaAs QW are with the regime of  values predicted by our theory,
but one order of magnitude smaller than those by the earlier model.
Moreover, it has been reported very recently that
there is a big discrepancy between the   earlier simplified treatment 
of the DP effect and the experiment on the SDT.
 It is shown that  the theoretical SDT is about one order 
of magnitude larger than the experiment data in $n$-type 
bulk GaAs for certain electron densities.\cite{song} 

It is seen that our many-body  result is better than the earlier 
simplified treatments both qualitatively and quantitatively. 
The reason that our model is more precise than the earlier one lies
on the fact  that the earlier model is based on the single particle 
picture  which does not count for the dephasing due to the 
inhomogeneous broadening inherited in the DP term,
which is the result of the many body effect.\cite{wu_pss_2000,%
wu_ssc_2002,wu_jpsj_2001,wu_epjb_2000,wu_js_2001} 
By comparing the theoretical SDT predicated by the two models, we can
see that the spin dephasing 
due to the inhomogeneous broadening is much more important. In the case we
calculated, the spin dephasing is dominated by the inhomogeneous
broadening. Therefore, it is easy to understand why the earlier
simplified treatment of the DP mechanism gives much slower
spin dephasing. 

The temperature dependence of the SDT can be 
understood once the spin dephasing due to the inhomogeneous broadening is
taken into account: When the temperature increases, the inhomogeneous
broadening is reduced as the electrons are distributed to the wider
$k$-states. As a result, the number of electron occupation 
on each ${\bf k}$ state is reduced. It is further
noted that this reduction is mild as a function of the
temperature. Therefore, the temperature dependence is quit mild unless
it is within the regime where the HF term plays an important role in
the spin dephasing.  

In the region where the HF term is important, in addition to the above
mentioned two effects of the temperature on the spin dephasing, 
the temperature dependence of the HF term should also be taken into
account. 
In general, the temperature dependence of the SDT due to the combination of
these three effects is too complicated to be described by a monotonic
function. We replot the SDT as a function of the
temperature in Fig.~\ref{fig6}(b), for a typical high polarization
$P=75$~\%. 
We can see that due to the reduction of
the HF term, the reduce of the detuning is removed and 
SDT drops dramatically with the increase of the temperature 
in the impurities free sample. 
While for the system with  impurity concentration
$N_i=0.1 N_e$, 
the HF term is not as important as in the impurities
free sample, and the SDT is insensitive to the temperature.
\begin{figure}[htb]
  \psfig{figure=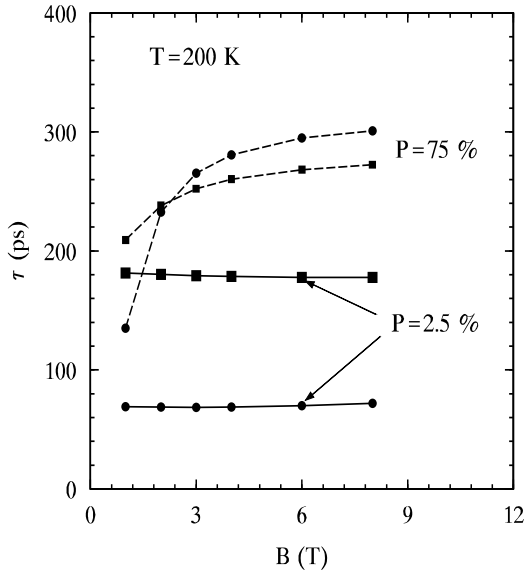,width=9.cm,height=8.5cm,angle=0}
  \caption{Spin dephasing time $\tau$ versus the applied magnetic
    field 
    for different spin polarizations and different
    impurity levels. 
    Solid curve with dots: $N_i=0$, $P=2.5\ \%$; 
    Solid curve with squares: $N_i=0.1N_e$, $P=2.5\ \%$; 
    Dashed curve with dots: $N_i=0$, $P=75\ \%$; 
    Dashed curve with squares: $N_i=0.1N_e$, $P=75\ \%$.}
  \label{fig7}
\end{figure}

\subsection{Magnetic field dependence of the spin dephasing}

We now investigate the magnetic field dependence of the spin dephasing. 
In Fig.~\ref{fig7}, we plot the SDT versus the applied magnetic field
for different impurity levels and different spin polarizations. 
It is seen from the figure that in small polarization regime 
for both impurity free and doped samples, the SDT
is almost a constant when the magnetic field varies from 1~T to 8~T. 
Whereas, in the high polarization
regime, the SDT increases with the magnetic field when the magnetic
field varies from 1~T to 4~T and then saturates when the magnetic field
is larger than 4~T. It is noted that the transverse magnetic field
imposes two effects on the spin dephasing. One is that 
in the presence of a magnetic field, the electron spins
undergo a Larmor precession around the magnetic field. This
precession  suppresses the precession about the effective magnetic field
${\bf h}({\bf k})$ of the DP term.\cite{meier,bronold} 
Therefore the SDT increases with the magnetic field. 
However, in the presence of the transverse magnetic field, the
spin precession frequency between the spin-up and -down band is 
$\sqrt{\bigl(g\mu_BB-h_x({\bf k})\bigr)^2+h_y^2({\bf k})}$. Hence 
the transverse magnetic field may introduce an additional inhomogeneous
broadening for the electrons and
consequently results in a shorter SDT. 
In the cases we study,  the SDT is almost unchanged with the magnetic
field  with the combined contributions from these two effects.

In additional to the above mentioned effects of the magnetic field on
the spin
dephasing, one can further see from Fig.~\ref{fig7}, that 
for large spin polarization, the magnetic
field enhances the SDT. As we mentioned before,
for large spin polarization, the contribution from the HF term is
important. Therefore, 
the magnetic field can further affect the spin dephasing rate through
the enhancement of the HF term in the high initial spin polarization
regime.  
It is shown in the figure that, for high initial spin polarization,
when the magnetic field increases, the  
enhanced HF term results in a fast decrease of spin dephasing
rate. Nevertheless, after the magnetic field is high enough, the HF
term arrives it 
maximum. As a result, the SDT saturates with the magnetic field. 

\begin{figure}[htbp]
  \psfig{figure=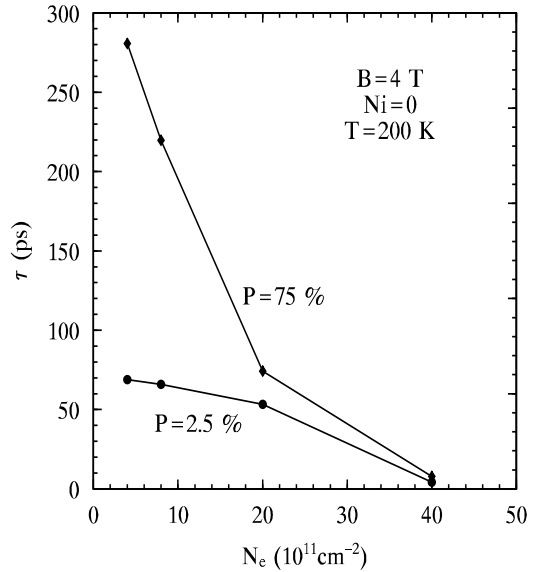,width=9.cm,height=8.5cm,angle=0}
  \caption{Spin dephasing time $\tau$ versus the total electron
    density $N_e$ for a GaAs QW with $T=200$~K, $B=4$~T, $N_i=0$ and
    $P=2.5$~\% ($\bullet$) and $P=75$~\% ($\blacklozenge$). 
    The lines are plotted for the aid of eyes.
  }
  \label{fig9}
\end{figure}

\subsection{Electron density dependence of spin dephasing}

We now turn to the electron density dependence of the spin
dephasing. In Fig.~\ref{fig9} we plot the SDT as a function of the
electron density for both low and high spin
polarizations.
It can be
seen from the figure that for low polarization,
the SDT decreases with electron density: from
$70$~ps for $N_e=4\times 10^{11}$~cm$^{-2}$ to $4$~ps for $N_e=40\times
10^{11}$~cm$^{-2}$. This is because with the increase of the electron
density, more electrons are distributed at large momentum states and
strengthen the DP effect as the DP term increases with the momentum. 
For the high spin polarization, the SDT also decreases with the
electron density, however, with a much faster speed. 
When the initial spin polarization
is 75\%, the SDT decreases from 280~ps for $N_e=4\times
10^{11}$~cm$^{-2}$ to 8~ps for $N_e=40\times 10^{11}$~cm$^{-2}$. This
is understood that when the electron density increases, the effect of HF
term becomes less important comparing to the increase of the DP term. 

\subsection{The effect of quantum well width on the spin dephasing}

\begin{figure}[htb]
  \psfig{figure=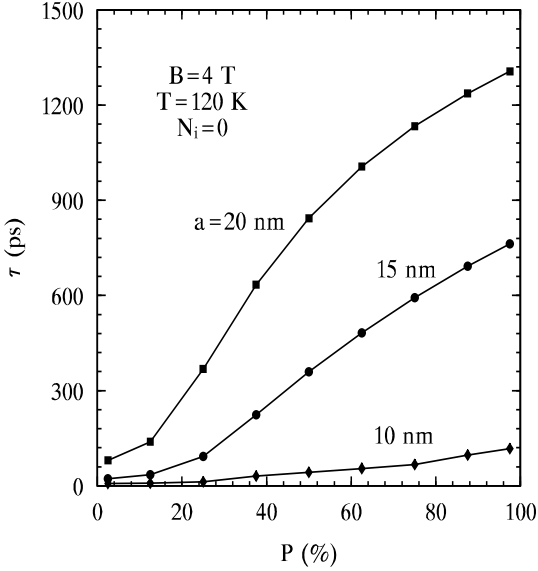,width=9.cm,height=8.5cm,angle=0}
  \caption{Spin dephasing time $\tau$ versus the initial spin
    polarization $P$ at $T=120$~K and $N_i=0$ in three quantum wells
    with different widths: 
    Diamond ($\blacklozenge$): $a=10$~nm; 
    Circle ($\bullet$): $a=15$~nm; 
    Square ($\blacksquare$): $a=20$~nm.
    The lines are plotted for the aid of eyes.
  }
  \label{fig10}
\end{figure}

\begin{figure}[htb]
  \psfig{figure=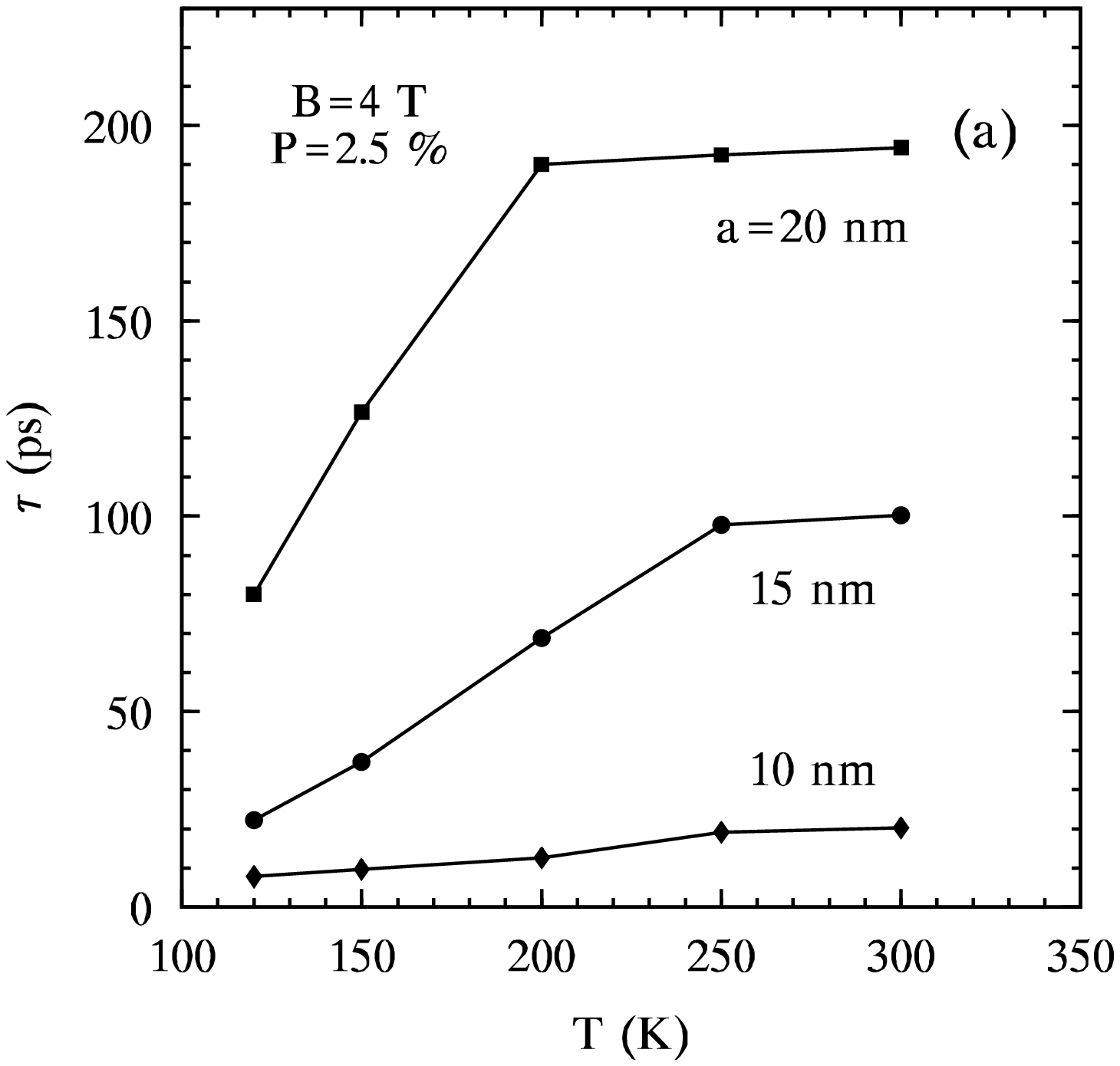,width=9.cm,height=8.5cm,angle=0}
  \psfig{figure=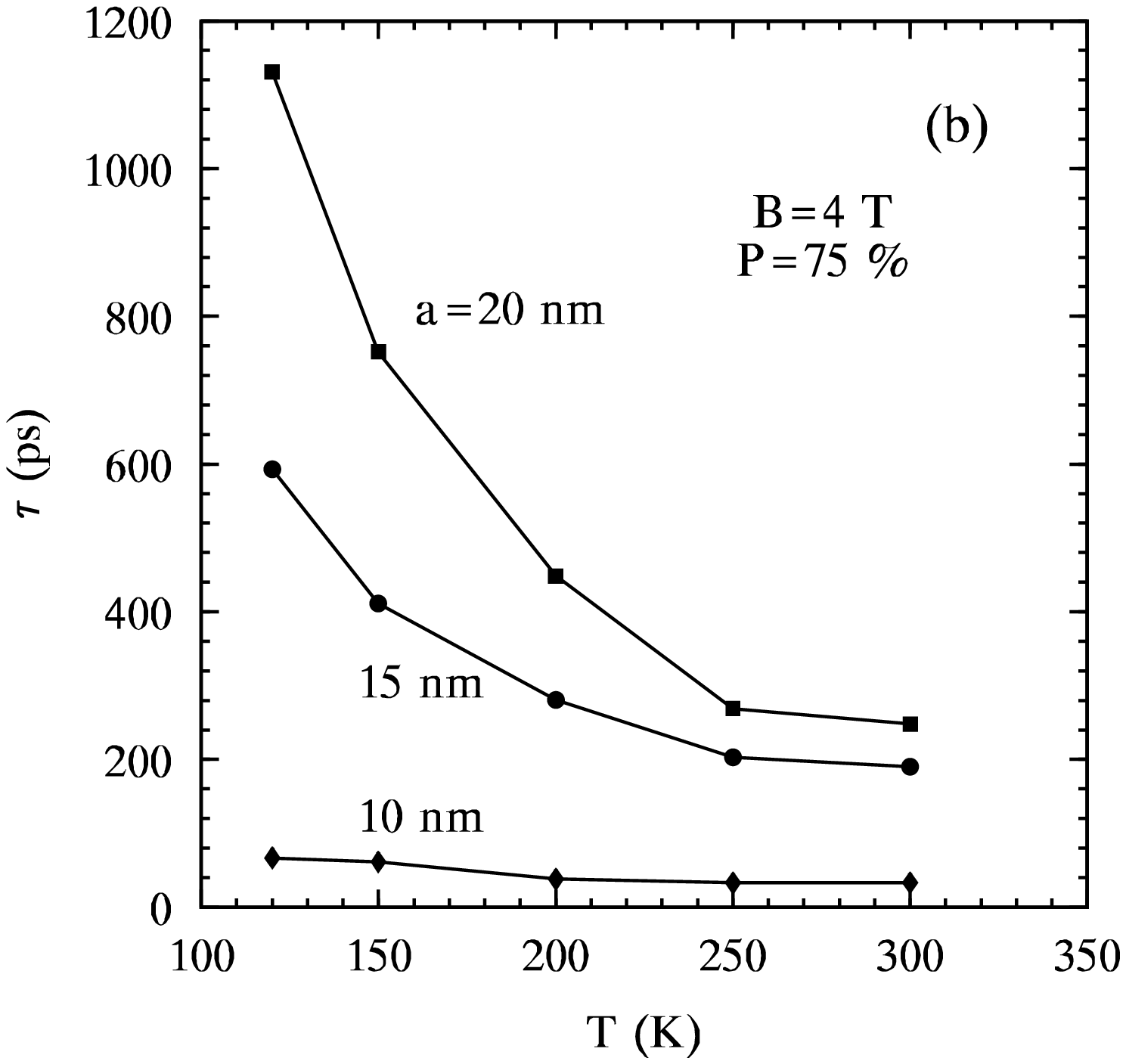,width=9.cm,height=8.5cm,angle=0}
  \caption{Spin dephasing time $\tau$ versus temperature $T$ 
    in three quantum wells with different widths: 
    Diamond ($\blacklozenge$): $a=10$~nm; 
    Circle ($\bullet$): $a=15$~nm; 
    Square ($\blacksquare$): $a=20$~nm.
    (a), the initial spin polarization $P=2.5$~\%; (b) $P=75$~\%. 
    The lines are plotted for the aid of eyes.
  }
  \label{fig11}
\end{figure}

It is noted from the experiments that the SDT of quantum wells is much
smaller than that of bulk material. This is because
the DP term in the quasi-two dimensional quantum well contains a term
which is proportional to the $\langle k_z^2\rangle=(\pi/a)^2$.
For a regular quantum well in the order of 10\ nm, this term is
much larger than the square of Fermi vector $k_F^2$, when the electron
density is up to the order of  
$10^{11}$\ cm$^{-2}$. Therefore the DP term in the quantum well is
greatly enhanced. Moreover, the smaller the width is, the larger
the DP term becomes and the faster  the SDT turns to be.
In this subsection we  study the effect of the quantum well width on
the spin dephasing. In Fig.~\ref{fig10} 
the SDT is plotted as a function 
of the initial spin polarization for three different widths. 
Moreover, in Figs.~\ref{fig11}(a) and (b) we plot the SDT as a function of the
temperature for two initial spin polarizations in three different
quantum wells. These figures 
clearly show that, for all situations, 
the spin dephasing rate decreases 
with the increase of the well width.
Especially when the width is decreased by only 50\ \% from 
20\ nm, the SDT is reduced by one order of magnitude. 

\section{Conclusion}

In conclusion, we have performed a systematic investigation of the DP
effect on the spin dephasing of $n$-typed GaAs QW's for high
temperatures under moderate
magnetic fields in Voigt configuration. Based on the nonequilibrium
Green's function theory, we derived a set of kinetic Bloch equations
for a two-spin-band model. This model includes the electron-phonon,
electron-impurity scattering as well as the electron-electron
interaction. By 
numerically solving the kinetic Bloch equations, we study the time
evolution of electron densities in each spin band and the spin
coherence -- the correlation between spin-up and -down bands. The spin
dephasing time is calculated from the slope of the envelope of the
time evolution of the incoherently summed spin coherence. We therefore
are able to study in detail how this dephasing time is affected by
spin polarization, temperature, impurity level, magnetic field and
electron density. Different from the earlier studies on spin dephasing
based on the single particle model which only considers the effective
SF scattering, our theory also takes account of the contribution of
many-body effect on the spin dephasing.\cite{wu_pss_2000,wu_ssc_2002,%
wu_jpsj_2001,wu_epjb_2000,wu_js_2001} In fact, for the $n$-typed 
semiconductors and the spin polarization studied in the experiments,
this many-body dephasing effect is even more important than
the effective SF scattering as it can be one order of magnitude 
larger than the later. Equally remarkable is that, as we include all
the scattering, especially the Coulomb scattering in our many-body
theory, now we are able to calculate the spin dephasing with extra
large (up to 100\ \%) initial spin polarization. 

It is discovered that the SDT increases dramatically with the initial
spin polarization. 
We stress here that the SDT is defined to be the inverse of the spin
dephasing rate, instead of the total life time which naturally
increases with the initial spin polarization. 
At low impurity level and for low temperature, the
magnitude of SDT increases more than one order when the initial
polarization goes from 0 to 100~\%. 
It is discovered that this
dramatic increase originates from the HF contribution of the
electron-electron Coulomb interaction. The HF term forms an effective
magnetic field which affects the spin dephasing. As the longitudinal
component of the HF term effectively reduces the ``detuning'' between
the spin-up and -down bands, the spin dephasing becomes much slower in
high polarization region. Due to the fact that the dramatic increase
comes from
the HF term, the magnitude of the increasement is therefore affected
by all the factors that influences the HF term, such as
temperature, impurity scattering, magnetic field as well as the
electron density. When the temperature or the impurity density
increases, the effective magnetic field formed by the HF term for a
given initial spin polarization decreases. Therefore, the increase
in $\tau$-$P$ curve is reduced, and the $\tau$-$P$ curve becomes
flatter. Nevertheless when the magnetic field increases, the HF term is
enhanced and the increasement in the $\tau$-$P$ is enhanced 
accordingly. When the magnetic field increases beyond 4~T, the HF
saturates. Consequently after 4~T, the
increase in $\tau$-$P$ curves only gets slight changes
with the magnetic field. 

For low spin polarized regime, the SDT increases with the temperature.
This is contrary to the result of earlier simplified
single-particle calculation where the SDT always decreases with the
increase of the temperature. Moreover, the SDT predicted by our many-body
calculation is one order of magnitude faster than the earlier result.
We show that our results are in agreement with the
experiments both qualitatively and quantitatively.
The physics of this feature is due to the additional 
many-body spin dephasing channel due to the 
inhomogeneous broadening provided by the  DP term, which by 
combining with the SC scattering also causes spin dephasing.
In the situation we studied, the spin dephasing is dominated by the
many-body dephasing effect.
With the increase of the temperature, the inhomogeneous broadening
reduces and the SDT increases. 

The effect of the electron-impurity scattering on the
spin dephasing is also studied. 
The SDT increases in the presence of the impurity scattering. 
This is because the impurity scattering 
redistributes the
electrons to an isotropic state which in turn reduces the spin
dephasing.  

In high spin polarization region, in additional to the above mentioned 
role the scattering plays in the spin
dephasing, the scattering  also affects  the HF term. This brings more
complication in the study of the spin dephasing. In general the SDT can
not be described by a monotonic function of the temperature or the
impurity concentration in high polarization region. For a typically
high initial spin polarization, $P=75$\ \%, 
the SDT decreases dramatically
with the temperature as the reduction of the detuning is suppressed
when the 
temperature increases. Whereas when the impurity concentration is $0.1
N_e$, the SDT is insensitive to the temperature.  

As the magnetic field causes the electron spins to precess about it, this
precession will suppress the precession about the effective magnetic
field ${\bf h}({\bf k})$ originated from the DP effect. However, 
the transverse magnetic field in the Voigt configuration also introduces
additional inhomogeneous broadening in the momentum space and enhances
the spin dephasing. 
As a result of the combining effects, for the condition we studied, 
the spin dephasing is almost 
unchanged with the magnetic field in low spin polarization region.
The magnetic field also enhances the HF term,
therefore the $\tau$-$B$ curve
gets a faster increase in the high polarization region when the
magnetic field is smaller than 4~T. In the region of the magnetic
field higher than 4~T, the HF term achieves its maximum. The effect
of the magnetic field on the HF term saturates, the SDT saturates
accordingly. 

Our calculation also shows that when then electron density increases,
the SDT decreases. This is because with the increase of the electron
density, more electrons are distributed at larger momentum states and
consequently the DP term is strengthened. While as the width of the
quantum well increases, the SDT decreases as the DP term reduces. 

In summary we have performed a thorough investigation of the spin
dephasing 
in $n$-typed GaAs QW's for high temperatures. Many new features which
have not been investigated both theoretically and experimentally 
are predicted in a wide range of
parameters.

\acknowledgments
MWW is supported by the  ``100 Person Project'' of Chinese Academy of
Sciences and Natural Science Foundation of China under Grant
No. 10247002.


\end {document}